\documentclass[12pt]{article}
\usepackage{latexsym,epsfig,graphicx,a4wide,amssymb,cite}

\def\a{\alpha}
\def\b{\beta}
\def\g{\gamma}
\def\G{\Gamma}
\def\d{\delta}
\def\D{\Delta}
\def\e{\eta}
{\rm }
\def\la{\lambda}

\def\k{\kappa}
\def\m{\mu}
\def\n{\nu}
\def\r{\rho}
\def\o{\omega}
\def\p{\pi}
\def\s{\sigma}
\def\S{\Sigma}
\def\t{\tau}

\def\th{\theta}
\def\z{\zeta}

\def\be{\begin{equation}}
\def\ee{\end{equation}}
\def\bea{\begin{eqnarray}}
\def\eea{\end{eqnarray}}
\def\nn{\nonumber}

\def\de{\partial}

\def\H{\mathcal{H}}

\def\O{\mathcal{O}}
\def\R{\mathcal{R}}

\def\2{\frac{1}{2}}
\def\4{\frac{1}{4}}

\catcode`\@=11

\@addtoreset{equation}{section}

\def\@normalsize{\@setsize\normalsize{15pt}\xiipt\@xiipt
\abovedisplayskip 14pt plus3pt minus3pt%
\belowdisplayskip \abovedisplayskip
\abovedisplayshortskip  \z@ plus3pt%
\belowdisplayshortskip  7pt plus3.5pt minus0pt}
\def\small{\@setsize\small{13.6pt}\xipt\@xipt
\abovedisplayskip 13pt plus3pt minus3pt%
\belowdisplayskip \abovedisplayskip
\abovedisplayshortskip  \z@ plus3pt%
\belowdisplayshortskip  7pt plus3.5pt minus0pt
\def\@listi{\parsep 4.5pt plus 2pt minus 1pt
            \itemsep \parsep
            \topsep 9pt plus 3pt minus 3pt}}

\def\underline#1{\relax\ifmmode\@@underline#1\else
        $\@@underline{\hbox{#1}}$\relax\fi}
\@twosidetrue \relax

\catcode`@=12

\catcode`\@=11

\def\section{\@startsection{section}{1}{\z@}{3.5ex plus 1ex minus
   .2ex}{2.3ex plus .2ex}{\large\bf}}

\def\ps@headings{\def\@oddfoot{}\def\@evenfoot{}
\def\@oddhead{\hbox{}\hfill
        \makebox[.5\textwidth]{\raggedright\ignorespaces --\thepage{}--
        \hfill }}
\def\@evenhead{\@oddhead}
\def\subsectionmark##1{\markboth{##1}{}}
}

\ps@headings

\catcode`\@=12

\def\a{\alpha}
\def\r{\rho}
\def\s{\sigma}
\def\t{\tau}
\def\m{\mu}
\def\n{\nu}
\def\k{\kappa}
\def\th{\theta}
\def\g{\gamma}\def\G{\Gamma}
\def\L{\Lambda}\def\l{\lambda}
\def\D{\Delta}
\def\la{\langle}
\def\ra{\rangle}
\def\o{\omega}\def\O{\Omega}
\def\d{\delta}
\def\p{\partial}

\def\z{\zeta}
\def\de{\partial}
\def\nn{\nonumber}

\def\half{\textstyle{\frac{1}{2}}}

\def\bdoc{\begin{document}}
\def\edoc{\end{document}}

\def\beq{\begin{equation}}
\def\eeq{\end{equation}}
\def\bea{\begin{eqnarray}}
\def\eea{\end{eqnarray}}
\def\ben{\begin{enumerate}}
\def\een{\end{enumerate}}
\def\la{\langle}
\def\ra{\rangle}
\def\a{\alpha}
\def\b{\beta}
\def\g{\gamma}\def\G{\Gamma}
\def\d{\delta}\def\D{\Delta}
\def\e{\epsilon}

\def\th{\theta}
\def\k{\kappa}
\def\l{\lambda}
\def\m{\mu}
\def\n{\nu}
\def\o{\omega}
\def\p{\pi}
\def\r{\rho}
\def\s{\sigma}
\def\t{\tau}
\def\L{{\mathcal L}}
\def\S{\Sigma }
\def\gsim{\; \raisebox{-.8ex}{$\stackrel{\textstyle >}{\sim}$}\;}
\def\lsim{\; \raisebox{-.8ex}{$\stackrel{\textstyle <}{\sim}$}\;}
\def\gtrsim{\gsim}
\def\lessim{\lsim}
\def\loc{{\rm local}}
\def\vm{v_{\rm max}}
\def\bh{\bar{h}}
\def\del{\partial}
\def\nab{\nabla}
\def\half{{\textstyle{\frac{1}{2}}}}
\def\fourth{{\textstyle{\frac{1}{4}}}}

\def\bD{{\bf D}}
\def\bE{{\bf E}}
\def\bF{{\bf F}}
\def\bB{{\bf B}}
\def\bP{{\bf P}}
\def\bV{{\bf v}}
\def\bv{{\bf v}}
\def\bx{{\bf x}}
\def\by{{\bf y}}
\def\bz{{\bf z}}
\def\ba{{\bf a}}
\def\bd{{\bf d}}
\def\bs{{\bf s}}
\def\bn{{\bf n}}
\def\bp{{\bf p}}

\def\O{\Omega}

\def\U{\bf V}
\def\br{{\bf r}}
\def\bnab{{\bf \nab}}

\def\tE{\tilde{E}}
\def\tL{\tilde{L}}

\newcommand{\scri}{\mathscr{I}}
\newcommand{\sun}{\ensuremath{\odot}}%
\def\e{{\mathrm e}}%
\def\g{{\mbox{\sl g}}}%
\def\Box{\nabla^2}%
\def\d{{\mathrm d}}%

\def\R{{\rm I\!R}}%
\def\ie{{\em i.e.\/}}%
\def\eg{{\em e.g.\/}}%
\def\etc{{\em etc.\/}}%
\def\etal{{\em et al.\/}}%
\def\HRULE{{\bigskip\hrule\bigskip}}

\def\d{{\mathrm{d}}}
\def\J{{\mathscr{J}}}
\def\L{{\mathscr{L}}}
\def\H{{\mathscr{H}}}
\def\T{{\mathscr{T}}}
\def\V{{\mathscr{V}}}

\def\sech{{\mathrm{sech}}}
\begin{document}

\begin{titlepage}
%
%


%

\begin{centering}
\vspace{1cm}
{\Large {\bf Massive Gravity with Anisotropic Scaling}}\\

\vspace{1.5cm}

 {\bf Bertha Cuadros-Melgar$^{1,2}$ $^{*}$}, {\bf Eleftherios Papantonopoulos}$^1$
 $^{**}$,\\
 \vspace{.1in}
 {\bf Minas~Tsoukalas}$^{1,3}$ $^{\flat}$ and {\bf  Vassilios Zamarias}$^1$ $^{\natural}$\\
 \vspace{.2in}
$^1$ Department of Physics, National Technical University of
Athens, \\
Zografou Campus GR 157 73, Athens, Greece. \\
\vspace{.1in}
 $^2$  Physics Department, University of Buenos Aires
FCEN-UBA and IFIBA-CONICET Ciudad Universitaria,
Pabell$\acute{o}$n 1, 1428, Buenos Aires,
Argentina.\\
\vspace{.1in}
 $^3$ Centro de Estudios Cient\'{\i}ficos (CECs), Casilla 1469,
 Valdivia, Chile.

\vspace{3mm}

\end{centering}
\vspace{2cm}

\begin{abstract}

We study a massive gravity theory which is Lorentz violating all
the way from ultraviolet to infrared energy scales.  At short
distances the theory breaks diffeomorphism invariance and time and
space scale differently. Dynamical metric fields are introduced
which upon linearization over a Minkowski background correspond to
Lorentz violating mass terms at large distances. We perform a
scalar perturbation analysis and we show that with an appropriate
choice of parameters the  theory is healthy without ghosts,
tachyons, strong coupling problems and instabilities.
\end{abstract}

\vspace{2.5cm}
\begin{flushleft}
$^{*}~~$ e-mail address: berthaki@gmail.com \\
$^{**} ~$ e-mail address: lpapa@central.ntua.gr \\
$ ^{\flat}~~$ e-mail address: minasts@central.ntua.gr\\
$ ^{\natural}~~$ e-mail address: zamarias@central.ntua.gr

\end{flushleft}
\end{titlepage}

\section{Introduction}

Einstein's theory of General Relativity (GR) upon linearization on
a Minkowski background, produces two physical massless propagating
spin-2 modes. These gravitational waves correspond to massless
particle excitations, the graviton, with two independent
polarization states, of helicity $\pm2$. Any attempt to give mass
to the graviton is problematic  at classical level and upon
quantization, confronts with  the severe short distance behaviour
of the graviton propagator, since quantum gravity is not
renormalizable.

Recently, there is a lot of interest in  deviating  from General
Relativity (GR) at ultra-large distance and time scales. The
motivation mainly comes from cosmology, and from the attempts to
explain the current observations that our universe accelerates.
 One way to produce models with infrared-modified (IR) gravity
is to give mass to the graviton.  This is a natural modification
of GR, since it amounts to simply giving a mass to the particle
which is already present in GR. The effect on the massive graviton
is to acquire at least three more degrees of freedom beyond the
two of the massless graviton.

A general approach to consider these models is to view them as
possible low energy limits of an unknown fundamental theory. This
stems from our lack of understanding of the behaviour of gravity
theory at short distances. In the quantum field theory language,
these models are described by an effective theory  above a certain
``ultraviolet'' scale $\Lambda_{UV}$.  This would mean that the
theory of gravity is a viable, phenomenologically acceptable
theory if all "unwanted" degrees of freedom decouple
 at energies above $\Lambda_{UV}$.

One of the first attempts to modify  GR was done by Fierz--Pauli
~\cite{FP}, introducing a mass term for the graviton.  A general
problem for such theories is that once  general coordinate
invariance is broken explicitly or spontaneously, gravity gets
modified, but new light degrees of freedom may appear among metric
perturbations, over and beyond spin-2 gravitons. These new degrees
of freedom may be ghosts or tachyons, which is often unacceptable.
In the Fierz--Pauli model the ghosts are absent, but in the limit
of vanishing  graviton mass $m$, the graviton propagator exhibits
the van~Dam--Veltman--Zakharov (vDVZ) discontinuity~\cite{VDV,Z}
originating from a scalar degree of freedom which does not
decouple in the massless limit. At the classical level the
Vainstein mechanism due to non-linearities makes this scalar
degree of freedom harmless
~\cite{Vainshtein:1972sx,Deffayet:2001uk}. However, at the quantum
level the theory becomes strongly coupled~\cite{AHGS} in the sense
that the new degree of freedom does not decouple above
$\Lambda_{UV}$ (for a review on massive gravity see
\cite{Hinterbichler:2011tt}).

To cure the undesirable features of the Fierz--Pauli model,  it
was proposed in \cite{Rubakov:2004eb} and further elaborated in
\cite{Rubakov:2008nh}, that the violation of Lorentz invariance
may give a healthy massive gravity theory in the infrared (IR)
\footnote{This model however does not address the Boulware-Deser
instability \cite{Boulware:1973my} which appears if any
nonlinearity is present in the theory. See \cite{deRham:2010kj}
for a suggested resolution of this problem.}. It was shown that by
considering a deformation of GR by Lorentz-violating graviton mass
terms, about flat spacetime and by setting  one of the masses
equal to zero and choosing appropriate conditions for the other
masses, the theory possesses desirable properties: no ghosts,  the
vDVZ discontinuity is absent and there is no  strong coupling
problem. A generalization of \cite{Rubakov:2004eb}, where all the
Lorentz violating masses are non-zero was discussed in
\cite{hep-th/0409124}.



A new theory of gravity was recently proposed \cite{Horava:2009uw}
which  breaks diffeomorphism invariance and it is supposed to be
an adequate UV completion of GR.  Its basic assumption is the
existence of a preferred foliation by three-dimensional constant
time hypersurfaces, which splits spacetime into space and time
\footnote{The behaviour of a higher-dimensional theory was
presented in \cite{Papantonopoulos:2011xa}, where the higher
dimensional diffeomorphism invariance was explicitly broken to its
foliation-preserving subgroup, leaving intact the four-dimensional
spacetime symmetry. See also \cite{He:2011hb}.}. This allows to
add higher order spatial derivatives of the metric to the action,
without introducing higher order time derivatives. This is
supposed to improve the UV behaviour of the graviton propagator
and render the theory power-counting renormalisable without
introducing ghost modes, which are common when adding higher order
curvature invariants to the action in a covariant manner
\cite{Stelle:1977ry}.

As in the Fierz--Pauli model where the breaking of general
coordinate invariance introduces new degrees of freedom, the
explicit  breaking of general covariance by the preferred
foliation of spacetime introduces a new scalar degree of freedom
in addition to the usual helicity-2 polarizations of the graviton.
 In the version with
``projectability'' and for maximally symmetric backgrounds, the
scalar is either (classically) unstable  or it becomes a ghost
(quantum-mechanically unstable)
  \cite{Sotiriou:2009bx,Charmousis:2009tc,Koyama:2009hc,Bogdanos:2009uj} \footnote{See however
\cite{Mukohyama:2010xz} and references therein for conditions
under which the  classical instability does not show up.}.

The non-projectable version of the theory suffers from strong
coupling at energies above a very low energy scale, and fast
instabilities. It was suggested in \cite{Blas:2009yd,Blas:2009qj}
that the extra mode may acquire a regular quadratic Lagrangian
with no ghosts and instabilities if the action
 is supplemented by certain type of Lorentz violating new terms, controlled by a dimensionless parameter
 $\eta$. If higher derivative terms are added in the theory, it was criticized in
 \cite{Papazoglou:2009fj}  that the strong problem
 persists in this extended Ho\u rava-Lifshitz theory, but it was claimed in \cite{Blas:2009ck} that there is
 a window of observationally accepted parameters, in which these terms are suppressed
 for a  certain UV scale. The strong coupling problem of the
 non-projectable version of the  Ho\u rava-Lifshitz theory was further discussed in
 \cite{arXiv:1003.5666, Padilla:2010ge} where it was shown that
 the strong coupling problem is not solved unless one introduces
 a low energy Lorentz violating scale. In section three we will
 discuss  extensively  the strong coupling problem which
 seems to be an endemic feature of modified gravity theories.

A central issue in all the theories that have been proposed to
describe gravity in the UV, is how to recover GR in the infrared.
This issue is not very well understood so far and little progress
has been done. All the couplings  which appear in front of kinetic
terms, interaction terms, high curvature terms are running
coupling constants which are scale dependent and supposed to run
through a renormalization group (RG) flow from UV to IR. In
three-dimensional TMG, the RG flow of the coupling of the
Chern-Simons term (which is responsible to give mass to the
three-dimensional graviton), the cosmological constant and
Newton's constant was studied \cite{Percacci:2010yk} \footnote{We
thank Ioannis Bakas for bringing this paper to our attention.}. It
was found that the coupling of the Chern-Simons term has vanishing
beta function \footnote{An Euclidean version of massive gravity
theory in the context of Ho${\rm\check{r}}$ava-Lifshitz theory was
discussed in \cite{Cai:2009ar}.}.

In 3+1 dimensions the Ho${\rm\check{r}}$ava-Lifshitz theory is
expected to have  a $z=3$ fixed point in the UV, where $z$
measures the degree of anisotropy between space and time, and to
flow to a $z=1$ fixed point in the IR, which is just the classical
Einstein-Hilbert gravity theory. This procedure is better
described by a dimensionless parameter $\lambda$ which appears in
the kinetic part of the action. At low energies, the resulting
action differs from that of GR only by the presence of this
parameter. Then it was conjectured that  the theory might have GR
as its low-energy limit, provided that $\lambda$ flows, by some RG
flow mechanism, to its GR value $\lambda=1$ in the infrared.

In this work we study a massive gravity theory which is Lorentz
violating all the way from UV to IR energy scales.  The theory in
short distances behaves like Ho\u rava: time and space scale
differently, and it is
 power-counting renormalizable. It also includes  a vector field,
 depending on space and time, of the type introduced in \cite{Blas:2009qj},
  which
 preserves a three-dimensional Euclidean symmetry.
In such a theory   the renormalization group flow  might not work
to the benefit of Lorentz invariance recovery allowing $\lambda$
to flow to $1$  and $\eta$ to flow to $0$. In this case the theory
would be Lorentz violating in the IR.

In such a theory it is natural to consider Lorentz violating mass
terms. We know that the introduction of Lorentz violating mass
terms in  the Fierz--Pauli theory improves its behaviour: there
are no ghosts and tachyons, the vDVZ discontinuity is absent and
the strong coupling scale  is high enough evading, the strong
coupling problem. Motivated by this, we introduce dynamical metric
fields which upon linearization over a Minkowski background,
correspond to Lorentz violating mass terms which however do not
violate the three-dimensional Euclidean space symmetry.  By
performing a scalar perturbation
 analysis we show that the presence of the mass terms extents the
parametric space of the
 theory, giving more freedom for the constaints to be satisfied as
 of not having ghosts,
 instabilities and tachyons. We also show that the strong coupling
 scale is high enough  as to avoid the strong coupling problem.

The work is organized as follows. In section 2 we introduce in the
extended Ho\u rava-Lifshitz theory Lorentz violating mass terms
and performing a scalar perturbation we calculate the quadratic
and cubic Lagrangians. In section 3 we discuss the effect of
non-zero masses to the theory and in section 4 we conclude.

\section{Extended Ho\u rava-Lifshitz Gravity}

 Consider the non-projectable version of the  Ho\u
rava-Lifshitz gravity
\begin{equation}\label{bpsaction}
S=\frac{M_{Pl} ^2}{2} \int d^3 x dt N\sqrt{g} \left( K^{ij}K_{ij}
- \lambda K^2 + R + \eta a_i a^i \right)\,,
\end{equation} where \be \label{ai}
a_i\equiv \frac{\partial_iN}{N}\; \ee is a three-vector introduced
in \cite{Blas:2009qj}, and $M_{Pl}$ is the Planck mass, $g$ is the
determinant of the spatial metric $g_{ij}$, $\lambda$ is a
dimensionless running coupling, and $\eta$ is a another running
coupling associated with the vector  $a_i$. This vector is
manifestly covariant under the Ho\u rava transformations.

We add in this action a certain combination of metric fields
which upon linearization over a flat background
\begin{equation}
g_{\mu\nu} = \eta_{\mu\nu} + h_{\mu\nu} \,
\end{equation}
take the form
\begin{equation}\label{massaction}
S_m = \frac{M_{Pl} ^2}{4} \int d^3 x dt N\sqrt{g} \left( c_0
h_{00} h^{00} - 2c_1  h_{0i}h^{0i} -c_2  h_{ij}h^{ij} +c_3  h_i ^i
h_j ^j + 2c_4  h_0 ^0 h_i ^i \right) \,.
\end{equation}
These terms respect the anisotropic scaling of the Ho\u rava
theory, between space and time [t]=-3 and [x]=-1,   provided that
the
 running couplings $c_i$ have dimension six. We expect these
 couplings, when the theory runs to the IR, to correspond to
 graviton masses of dimension two. The action
 (\ref{massaction})   with the identification $c_i=m_i^2$, corresponds to the
 Lorentz violating massive gravity in the IR, proposed in
 ~\cite{Rubakov:2004eb}, while for $c_0=0, c_i=m^2$ for
 $i=1,2,3,4$ corresponds to the Fierz--Pauli massive gravity
  \cite{FP}. Note that these terms respect a three-dimensional
  Euclidean space symmetry.

We are interested in  the behaviour of the scalar mode of the
graviton. Thus we consider the following metric ansatz
\be\label{perts1}
 N=e^{\a(t,x)}, \  \  \  N_i=\de_i \b(t,x),  \  \  \   g_{ij}=e^{2\z(t,x)} \delta_{ij}~, \label{gauge}
 \ee
 which differs from the most general scalar perturbation possibly  by a perturbation of $g_{\m\n}$ of the form $2 \de_\m \de_\n E$,
   which however can be  gauged away (see \cite{Papazoglou:2009fj}).

We are mainly concerned for the behaviour of the theory described
by the actions (\ref{bpsaction}) and (\ref{massaction}) in the IR.
  The perturbative analysis of the action (\ref{bpsaction}) has been carried out and we briefly   review it here.
    Since $N$ is spacetime dependent,
   both $\alpha$ and $\beta$ are spacetime dependent, but nevertheless they are
   not propagating dynamical fields  and they can be eliminated from the
   action. Then it is found that the quadratic action  for the physical mode $\z$ is
   \beq S_2=- \int  \d^3x \d t  \left[{1
\over c_\z^2}\dot{\z}^2 - {\eta- 2 \over \eta }(\de \z)^2  \right
]~, \label{quad} \eeq where \beq c_\z^2={1-\lambda \over 3\lambda
-1}~.\label{blasaction} \eeq

This is the action obtained in ~\cite{Blas:2009qj}, without the
higher order operators. If \beq \label{constr} c_\z^2<0 \ \ \ \
{\rm and} \ \ \ \ 0<\eta<2 \label{condition} \eeq there are no
ghosts or instabilities. The next order in  the perturbation
analysis leads to the cubic action \bea
\!\!\!S_3\!&=&\!\int \d t \d^3x \left\{\left( 1+\frac{4}{ \eta^2}  \right) \z (\de \z)^2
 - {2 \over c_\z^4} \dot{\z} \de_i \z {\de^i \over \D} \dot{\z}     \right. \nn \\
&& \left.~~~~~~~~~~~~~~~~~~~ +  \left(\frac{3}{2}+ {1 \over \eta}
\right)\! \left[{1 \over  c_\z^4}    \z \left( {\de_i \de_j \over
\D}\dot{\z}\right)^2 \!- {(2 c_\z^2+1)  \over  c_\z^4}  \z
\dot{\z}^2   \right]
  \right\} \label{cubic}.
  \eea

One can see that after  performing canonical normalization of the
kinetic term in the quadratic action, eq.~(\ref{quad}), as $\z=
|c_\z| \hat{\z}/M_{\rm pl}$, there is a strong coupling problem
for $c_\z \to 0$ ($\l \to 1$), {\it i.e.} the $\z$-interactions
become strong for energies above the scale $|c_\z| M_{\rm pl}$
\cite{Papazoglou:2009fj}.

Calculating the dispersion relation of the scalar gravitational
mode $\zeta$ it can be seen that, depending on the values of
$\lambda$ and $\eta$, the propagating velocity of this mode could
be larger than the gravitational helicity-2 modes
~\cite{Blas:2009qj}. This implies that at low energies Lorentz
invariance is broken. Therefore, any mass term added in the action
(\ref{bpsaction}), like the Fierz--Pauli mass term, should be
Lorentz violating at low energies.

To analyse the action (\ref{massaction}) under the scalar
perturbations (\ref{gauge}) we note
\begin{eqnarray}\label{perts}
h_{00} &=& h^{00} = -h_0 ^0 = -2\alpha \,,\nonumber \\
h_{0i} &=& -h_i ^0 = \partial_i \beta \,, \nonumber \\
h_0 ^i &=& -h^{0i} = \partial^i \beta \,,\nonumber \\
h_{ij} &=& 2\zeta \delta_{ij} \,,\,\,\, h^{ij} = 2\zeta
\delta^{ij}\,,\,\,\, h_i ^j = 2\zeta \delta_i ^j  \,.
\end{eqnarray}

Substituting Eqs.(\ref{perts}) into the total action $S + S_m$,
given in (\ref{bpsaction}) and (\ref{massaction}),  and after
appropriate partial integrations we obtain the following quadratic
Lagrangian
\begin{eqnarray}\label{lag2}
{\cal L}_2 &=& \frac{3}{2} (1-3\lambda) \dot\zeta^2 - (1-3\lambda)\dot\zeta\,\Delta\beta +
\frac{1}{2} (1-\lambda)(\Delta\beta)^2 + (\partial_i \zeta)^2 - 2\alpha\,\Delta\zeta - \frac{\eta}{2} \alpha\,\Delta\alpha + \nonumber \\
&& +m_0 ^2 \alpha^2 - \frac{1}{2} m_1 ^2 \beta\,\Delta\beta -
3(m_2 ^2-3m_3 ^2)\zeta^2 + 6m_4 ^2 \alpha\zeta\,.
\end{eqnarray}
where we have changed notation of $c_i$ to $m_i^2$ to indicate
that they have the dimension mass-squared.

Varying the quadratic action with respect to $\beta$ and $\alpha$
we obtain the momentum and Hamiltonian constraints as follows,
\begin{eqnarray}
\beta &=& \frac{(1-3\lambda)}{(1-\lambda)\Delta -m_1 ^2} \dot\zeta \,, \label{momentum}\\
\alpha &=& \frac{2(\Delta-3m_4 ^2)}{-\eta\Delta+2m_0 ^2}\zeta \,.
\label{hamiltonian}
\end{eqnarray}

Substituting these constraints back into (\ref{lag2}) and after
some partial integrations we get
\begin{eqnarray}\label{quadlag}
{\cal L}_2 &=& \dot\zeta \left[ \frac{1}{2} (1-3\lambda) \left( 3-\frac{(1-3\lambda)\Delta}{(1-\lambda)\Delta -m_1 ^2}
\right)\right] \dot\zeta - \nonumber \\
&&-\zeta \left[ \Delta + \frac{2(\Delta-3 m_4
^2)^2}{-\eta\Delta+2m_0 ^2}\right] \zeta - 3(m_2 ^2 -3
m_3^2)\zeta^2\,.
\end{eqnarray}
The equation of motion resulting from the above Lagrangian is
 \be
- \frac{1}{2}\,\frac{(1-3\lambda)(2 \Delta -3m_1
^2)}{(1-\lambda)\Delta-m_1 ^2}\ddot{\zeta} -\Big{\{}\left[ \Delta
+ \frac{2(\Delta-3 m_4 ^2)^2}{-\eta\Delta+2m_0 ^2}\right]  + 3(m_2
^2 -3 m_3^2)\Big{\}}\zeta=0~. \label{eqmotion} \ee To find the
spectrum  we canonically normalize the $\zeta$ field \be
\zeta=\frac{\mid c_{\zeta}\mid}{m_1 M_{Pl}} \left(\frac{\frac{2
\Delta}{m_1^2}-3}{\Delta-\frac{m_1^2}{1-\lambda}}\right)
^{-1/2}\zeta^{c}~, \label{normalized} \ee and we substitute it
back to (\ref{quadlag}). In Fourier space for $\zeta\propto
e^{-i\omega t+i\mathbf{p}\mathbf{x}}$ the dispersion relation
reads
\begin{equation}\label{dispersion}
\omega^2 = \frac{Q(p^2)}{P(p^2)} + \frac{3 (m_2 ^2 - 3m_3
^2)}{P(p^2)}\,,
\end{equation}
where ${\bf p}$ is the three-momentum, and the polynomials $P$ and
$Q$ are given by
\begin{eqnarray}
P(p^2) &=& \frac{1}{2}\,\frac{(1-3\lambda)(2 p^2 +3m_1 ^2)}{(1-\lambda)p^2+m_1 ^2}\,,\label{P}\\
Q(p^2) &=& \frac{(2-\eta)p^4 -2(m_0 ^2-6m_4 ^2)p^2 +18m_4 ^4}{\eta
p^2 +2m_0 ^2}\,.\label{Q}
\end{eqnarray}
From the dispersion relation (\ref{dispersion}) the rest mass of
the field $\zeta$ turns out to be
\begin{equation}\label{massz}
M_\zeta ^2= \frac{2}{1-3\lambda}\left( 3\frac{m_4 ^4}{m_0 ^2}+m_2
^2 -3m_3 ^2 \right)\,.
\end{equation}
Note that if $m_1^2=0$ then from (\ref{quadlag}) we see that there
are no ghosts in the theory provided that $c_\zeta ^2<0$. Also if
all the masses are equal to zero from the dispersion relation
(\ref{dispersion}) we see that we do not have any instabilities if
also $0<\eta<2 $. Therefore all the previous results are
reproduced in the limit of zero masses. We will discuss the
general case in the next section.

Subsequently, we calculate the next order in perturbation
analysis.  In this case it is necessary to consider cubic-type
contributions of $h_{\mu\nu}$ to the action. In particular, the
$h_{0i}$ components related to $\beta$ become important since they
are related to time derivatives of $\zeta$ whose interactions
exhibit strong coupling. Accordingly, we consider the following
combination,
\begin{equation}
{\cal L}_m ^c = N \sqrt{g} \left( \xi_1 h_{0i}h^{0i} h_0 ^0 + \xi_2  h_{0i}h^{0i} h_k ^k + \xi_3 h_{ij} h^{0j} h_0 ^i \right)\,
\end{equation}
with $\xi_i$ constants. Using Eqs.(\ref{perts}) this contribution gives,
\begin{equation}\label{cubich}
{\cal L}_m ^c = \rho_1 \alpha (\partial_i\beta)^2 + \rho_2 \zeta (\partial_i\beta)^2\,,
\end{equation}
where $\rho_1=-2\xi_1$ and $\rho_2=-2(3\xi_2+\xi_3)$.

Thus, the total cubic Lagrangian coming from Eqs.(\ref{bpsaction}), (\ref{massaction}), and (\ref{cubich}) reads
\begin{eqnarray}\label{lag3}
{\cal L}_3 &=& \frac{3}{2} (3\lambda-1)(\alpha-3\zeta)\dot\zeta^2 - (3\lambda-1)(\alpha-\zeta)\dot\zeta\Delta\beta - \frac{1}{2}(\alpha-3\zeta)(\partial_i\partial_j \beta)^2- \nonumber \\
&&- 2\Delta\beta \partial_i\zeta\partial^i \beta+ \frac{1}{2} \left[ \lambda\alpha +(\lambda-4)\zeta\right] (\Delta\beta)^2 -\frac{\Delta\alpha}{2}\zeta^2 -\alpha\zeta\Delta\zeta - \frac{\zeta^2}{2}\Delta\zeta - \alpha^2 \Delta\zeta +\nonumber\\
&&+ \frac{\eta}{2}(\alpha +3\zeta)(\partial_i \alpha)^2 + m_0 ^2 \alpha^3 + 3(m_0 ^2 +2m_4 ^2)\alpha^2\zeta - 3(m_2 ^2-3m_3 ^2 -6m_4 ^2)\alpha\zeta^2 +\nonumber\\
&&+\left( \frac{m_1 ^2}{2}+\rho_1\right) \alpha
(\partial_i\beta)^2 + \left(\frac{m_1 ^2}{2}+\rho_2\right) \zeta
(\partial_i\beta)^2 - m_1 ^2 \zeta\beta\Delta\beta - 9(m_2 ^2
-3m_3 ^2) \zeta^3\,.\nonumber \\
\end{eqnarray}
Using the momentum and hamiltonian constraints (\ref{momentum})
and (\ref{hamiltonian}) we finally obtain the cubic Lagrangian
\begin{eqnarray}\label{cubiclag}
{\cal L}_3 &=& 3(3\lambda-1) \left[ \frac{(\Delta-3m_4
^2)}{-\eta\Delta+2m_0 ^2}\zeta -\frac{3}{2}\zeta\right]\dot\zeta^2
+2\dot\zeta \left[\frac{(\Delta-3m_4 ^2)}{-\eta\Delta+2m_0 ^2}\zeta -\frac{\zeta}{2}\right]\left[\frac{(3\lambda -1)^2\Delta}{(1-\lambda)
\Delta-m_1 ^2}\dot\zeta\right] \nonumber \\
&-&\left[\frac{(\Delta-3m_4 ^2)}{-\eta\Delta+2m_0 ^2}\zeta
-\frac{3}{2}\zeta\right]
\left[\frac{(3\lambda-1)\partial_i\partial_j}{(1-\lambda)
\Delta-m_1 ^2}\dot\zeta\right]^2 -
 2\partial_i\zeta\left[\frac{(3\lambda-1)\Delta}{(1-\lambda)\Delta-m_1 ^2}\dot\zeta\right]\left[\frac{(3\lambda-1)
 \partial^i}{(1-\lambda)\Delta-m_1 ^2}\dot\zeta\right] \nonumber \\
&+&\left[\frac{\lambda(\Delta-3m_4 ^2)}{-\eta\Delta+2m_0 ^2}\zeta
+
\left(\frac{\lambda-4}{2}\right)\zeta\right]\left[\frac{(3\lambda-1)
\Delta}{(1-\lambda)\Delta-m_1 ^2}\dot\zeta\right]^2 -  \zeta^2
\left[\frac{\Delta(\Delta-3m_4 ^2)}{-\eta\Delta+2m_0
^2}\zeta\right] \nonumber \\
 &-& 2\zeta\Delta\zeta \left[\frac{(\Delta-3m_4 ^2)}{-\eta\Delta+2m_0 ^2}\zeta\right] - \frac{\zeta^2}{2}\Delta\zeta +4\eta\left[\frac{(\Delta-3m_4 ^2)}{-\eta\Delta+2m_0 ^2}\zeta
+\frac{3}{2}\zeta\right]\left[\frac{(\Delta-3m_4 ^2)\partial_i}{-\eta\Delta+2m_0 ^2}\zeta\right]^2  \nonumber \\
&-& 4\Delta\zeta \left[\frac{(\Delta-3m_4 ^2)}{-\eta\Delta+2m_0
^2}\zeta\right]^2 + 8m_0 ^2\left[\frac{(\Delta-3m_4
^2)}{-\eta\Delta+2m_0 ^2}\zeta \right]^3 +12(m_0 ^2 +2m_4 ^2)\zeta
\left[\frac{(\Delta-3m_4 ^2)}{-\eta\Delta+2m_0 ^2}\zeta \right]^2  \nonumber \\
&-& 6(m_2 ^2-3m_3 ^2-6m_4 ^2)\zeta^2\left[\frac{(\Delta-3m_4
^2)}{-\eta\Delta+2m_0 ^2}\zeta \right]
+\left(\frac{m_1 ^2}{2}+\rho_2\right)\zeta \left[\frac{(3\lambda-1)\partial_i}{(1-\lambda)\Delta-m_1 ^2}\dot\zeta\right]^2 \nonumber \\
&+& (m_1 ^2 +2\rho_1)\left[\frac{(\Delta-3m_4
^2)}{-\eta\Delta+2m_0 ^2}\zeta
\right]\left[\frac{(3\lambda-1)\partial_i}
{(1-\lambda)\Delta-m_1 ^2}\dot\zeta\right]^2 + \nonumber \\
 \nonumber \\
&-& m_1 ^2 \zeta \left[\frac{(3\lambda-1)}{(1-\lambda)\Delta-m_1
^2}\dot\zeta\right]\left[\frac{(3\lambda-1)\Delta}{(1-\lambda)
\Delta-m_1 ^2}\dot\zeta\right] - 9(m_2 ^2 -3m_3 ^2)\zeta^3\,.
\end{eqnarray}

We can note that taking in equation (\ref{cubiclag})  the zero
 mass  limit (also the $\xi$ couplings  tend
to zero), we recover the results of \cite{Papazoglou:2009fj} for
the cubic Lagrangian.

\section{The Effect of Non-zero Masses}

In this section we study the effect of the non-zero Lorentz
violating masses to the extended Ho\u rava-Lifshitz gravity
theory. In particular we find the general conditions the
parameters have to satisfy such that the scalar mode of the
graviton displays a healthy behaviour at large distances
\footnote{In \cite{Wang:2010uga} a mass term for the scalar
graviton was introduced in order to stabilize the Minkowski
background.}.

\subsection{Absence of Ghosts and Instabilities}

As we discussed in the previous section, at quadratic order in the
extended Ho\u rava-Lifshitz theory, the demand of not having
ghosts or instabilities imposes the condition (\ref{condition}) on
the parameters of the theory. In our case the quadratic Lagrangian
is modified with the presence of the mass parameters.

\begin{figure}[h!]
\centering
\includegraphics[width=14.0cm]{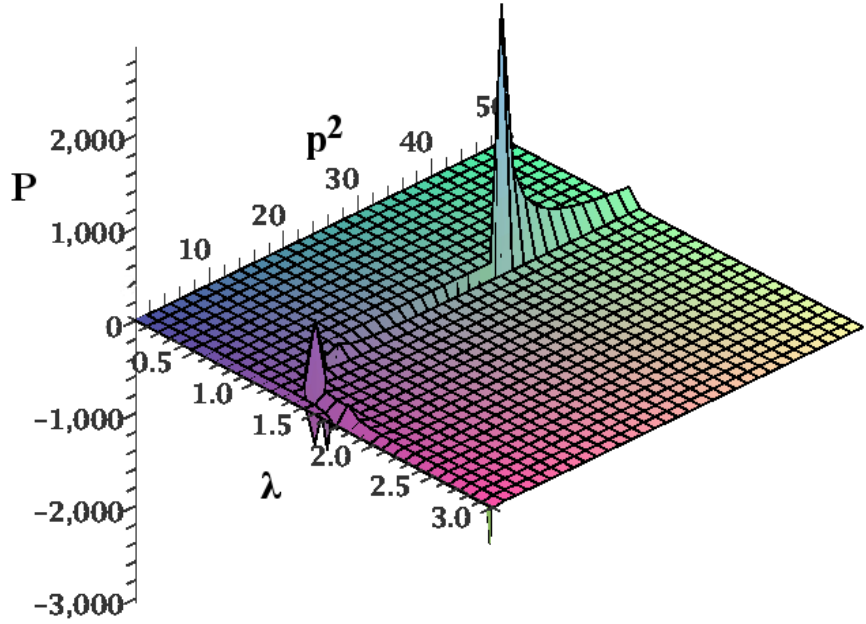}
\caption{$P(p^2)$ as a function of $p^2$ and $\lambda$.}
\label{3dP}
\end{figure}

\begin{figure}[h!]
\centering
\includegraphics[width=8.0cm]{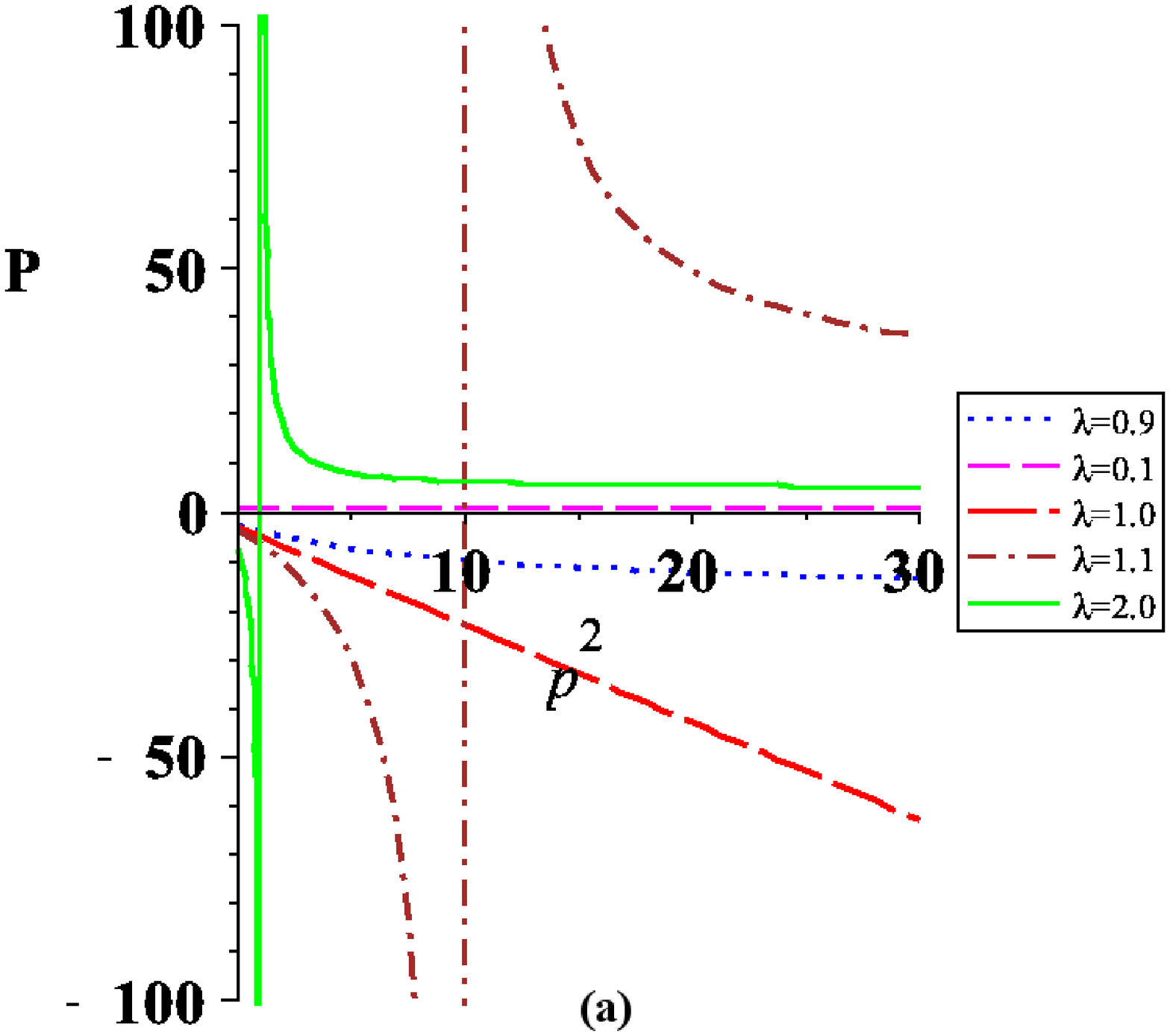}
\includegraphics[width=8.0cm]{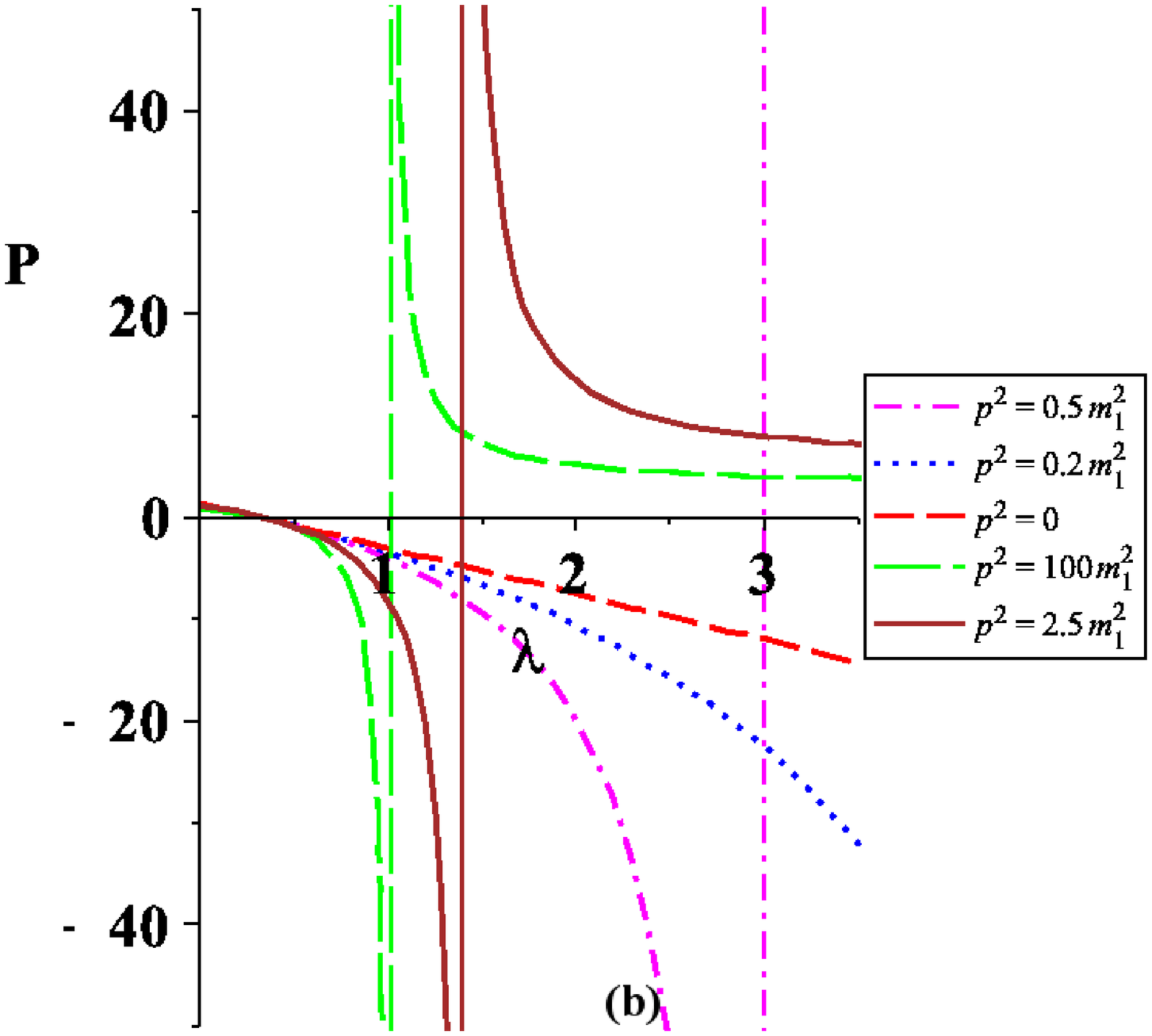}
\caption{$P(p^2)$ as a function of (a) $p^2$ and (b) $\lambda$. In
the first case the horizontal axis is in $m_1 ^2$ units, and in
the second case, the values of $p^2$ are multiples of $m_1 ^2$.}
\label{P(p)}
\end{figure}

In the kinetic part of the Lagrangian (\ref{quadlag}) the mass
parameter $m_1^2$ appears which imposes on the theory a new energy
scale. At momentum space the kinetic part is controlled by the
function $P(p^2)$ given by (\ref{P}). In order to avoid a ghost
field, $P(p^2)$ needs to be positive. In Fig.\ref{3dP} we show a
three-dimensional plot of $P$ as a function of $\lambda$ and
$p^2$, where $p^2$ is the three-dimensional spatial momentum
measured in units of $m_1 ^2$. We see that around $\lambda=1$ the
kinetic term can become negative. To have a better understanding
of the specific regions of the graph we plotted $P$ as a function
of each of the parameters in Fig.\ref{P(p)}.

A careful study of Fig.\ref{P(p)}(a) shows that when
$\frac{1}{3}<\lambda < 1$, $P$ is always negative for any value of
the 3-momentum. If $\lambda<\frac{1}{3}$, $P$ is always positive.
Whereas for $\lambda>1$, $P$ becomes positive after some high
value of momenta. As $\lambda$ grows, the region of positivity of
$P$ expands such that $P$ remains positive for $p^2 \geq m_1 ^2$.
This can also be noticed in Fig.\ref{P(p)}(b), where we can also
see that as the three-momentum grows, positive $P$ ends covering
all the region $\lambda> 1$.

Our theory is Lorentz violating at large distances. The important
issue is, however, how far from $\lambda=1$ it is allowed to go
taking under consideration the experimental constraints. This has
been discussed  recently in the literature (see the discussion in
\cite{Papazoglou:2009fj} and references therein). One constraint
 comes from the discrepancy between the effective
gravitational constant as measured in a gravitational experiment
$G_{\rm eff}$ and that measured in cosmology, $G_{\rm cosmo}$
\cite{Kehagias:2009is},\cite{Blas:2009qj}. This gives the maximum
upper bound for $\lambda$ as \beq \label{bound}
0<\lambda-1\lesssim  0.1\, . \eeq In fact, this is a mild
constraint since other constraints coming from Lorentz violations
would be much more stringent, requiring $\lambda$ to be even
closer to $1$.

Another important issue is how to constraint the mass scale
$m_1^2$. In general the presence of the Lorentz violating masses
introduces in the theory an  energy scale $E=\sqrt{mM_{Pl}}$
\cite{Rubakov:2004eb,hep-th/0409124}. An estimation of the mass
scale $m^2$ can be obtained from the experimental constants on the
graviton mass assuming that the graviton was created in the early
universe. As it was discussed in \cite{Rubakov:2008nh} an
estimation can be given from the observation of the slowdown of
orbital motion in binary systems. The agreement of these
observations with General Relativity implies that the mass of the
graviton cannot be larger than the characteristic  frequency of
the emitted gravitational waves. This gives \be \frac{m_G}{2\pi}
\equiv \nu_G \lesssim 3.10^{-5} Hz \thickapprox (10^{15}cm)^{-1}
\ee which on cosmological grounds is a large mass.

In our case Fig.\ref{P(p)} shows that  for values of $\lambda$
close to $1$, if the spatial momentum is well above the mass scale
$m_1^2$  we do not have ghosts.

\begin{figure}[h!]
\centering
\includegraphics[width=10.0cm]{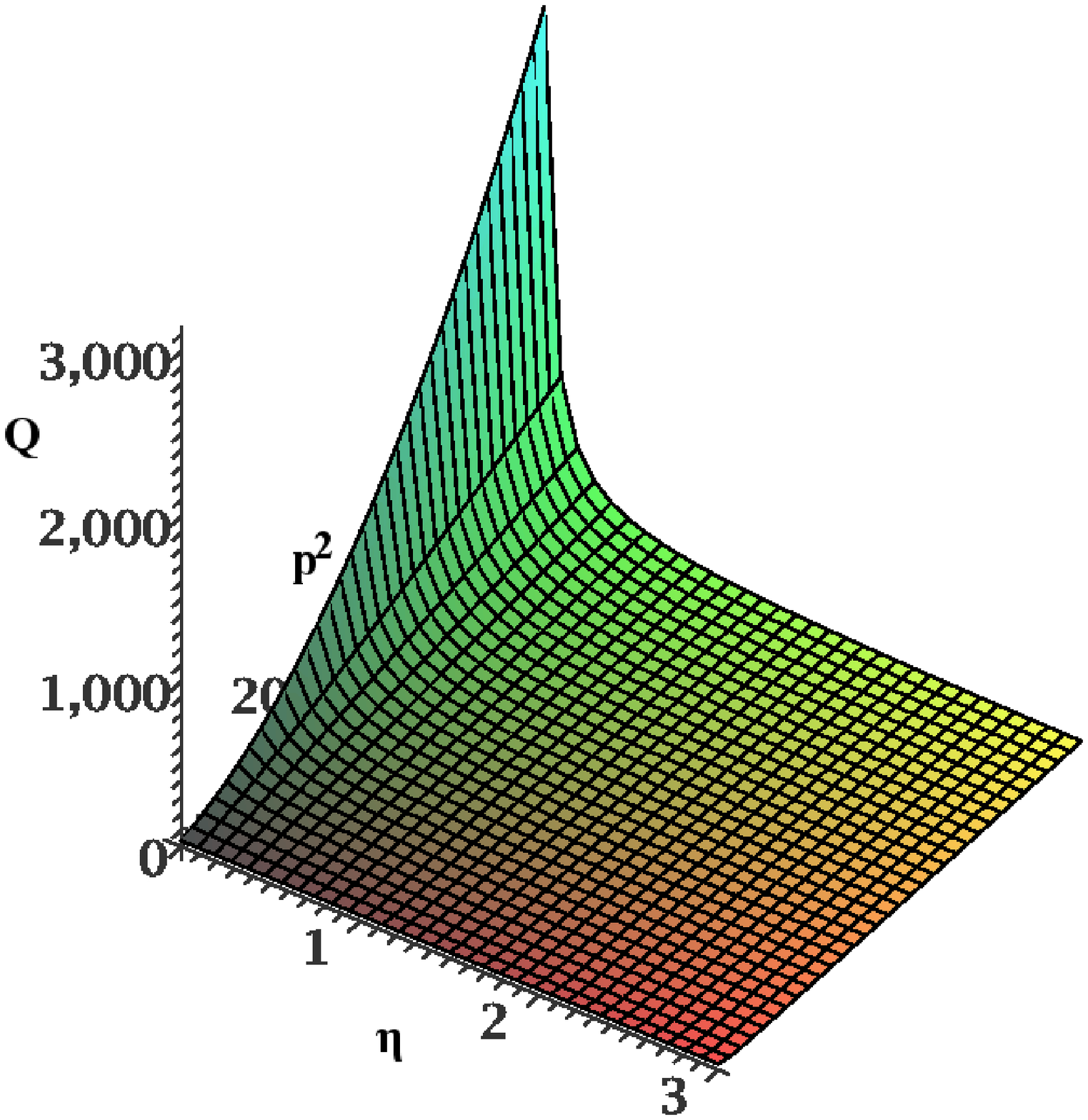}
\caption{$Q$ as a function of $\eta$ and $p^2$ for $m_4=2m_0$.}
\label{Q1}
\end{figure}

In the spatial part of the Lagrangian (\ref{quadlag}) other  mass
parameters also appear. These terms can give growing modes leading
to instabilities of the theory. When deriving the equation of
motion for the $\zeta$ field from the quadratic Lagrangian
(\ref{quadlag}),  the frequency $\omega^2$, having the dispersion
relation (\ref{dispersion}) in momentum space, appears in front of
the $\zeta$ term; therefore, to have a stable scalar mode
$\omega^2$ should be positive. We will investigate under which
conditions the $Q$ operator (\ref{Q}) is positive. In Fig.\ref{Q1}
we plotted $Q$ as a function of $\eta$ and $p^2$, having fixed the
parameters $m_0$ and $m_4$.

The effect of the relation between $m_4 ^2$ and $m_0 ^2$ is the
following.  When $m_0 ^2$ becomes much greater than $m_4 ^2$, Q can
become negative in some region that enlarges as the ratio
$\frac{m_0 ^2}{m_4 ^2}$ grows.

\begin{figure}[h!]
\centering
\includegraphics[width=8.0cm]{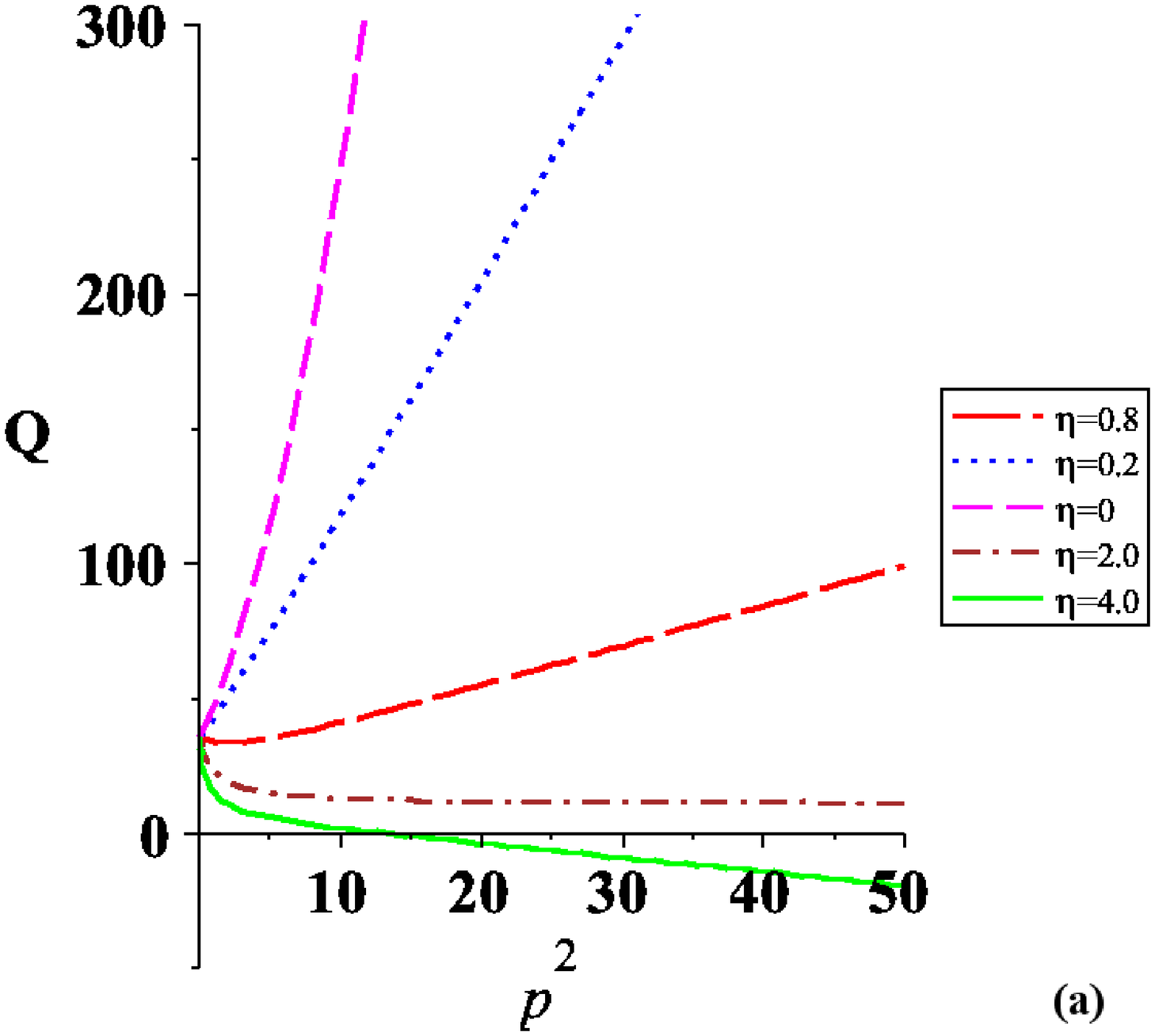}
\includegraphics[width=8.0cm]{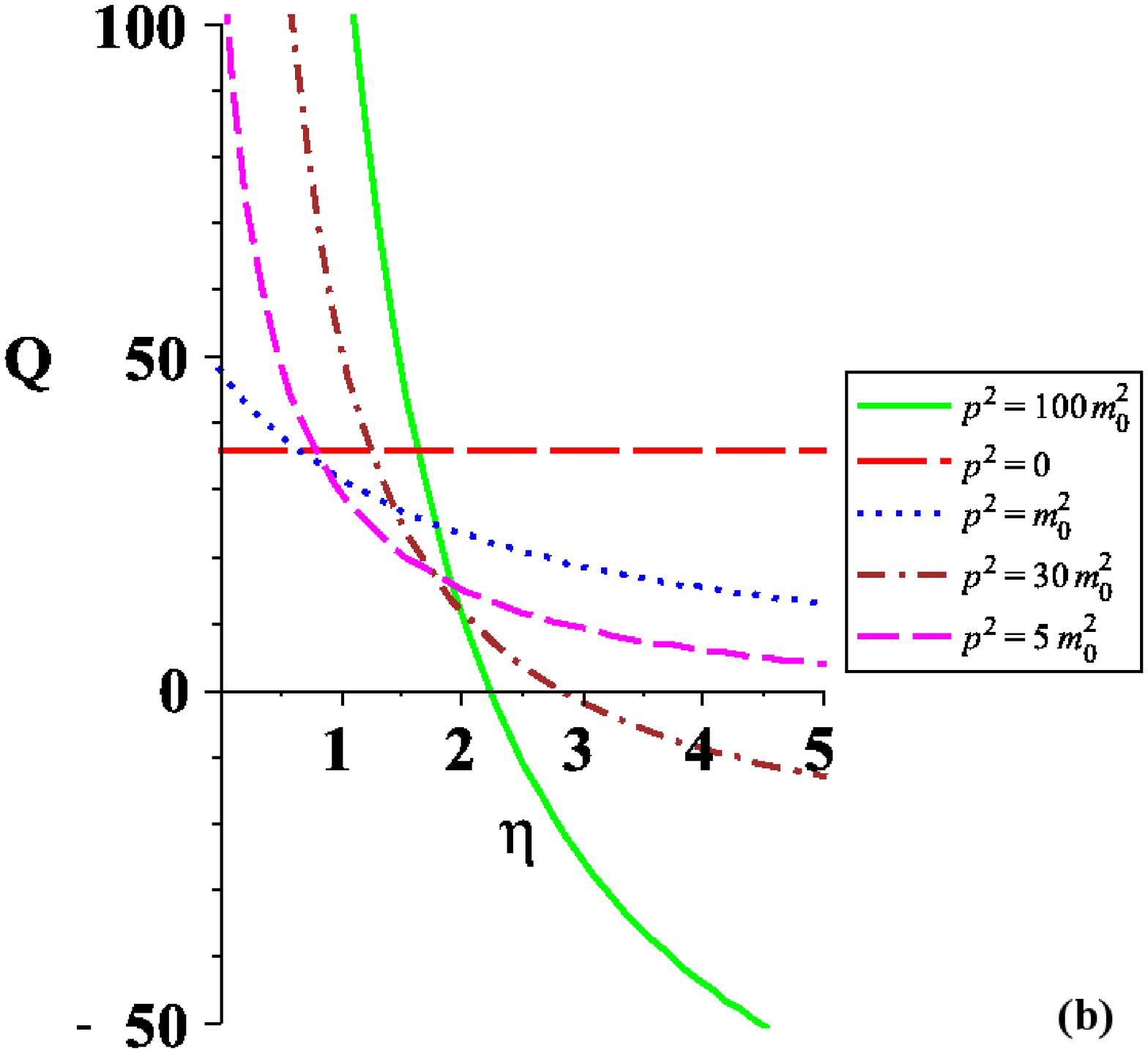}
\caption{$Q(p^2)$ as a function of (a) $p^2$ and (b) $\eta$. In the first case the horizontal axis is in $m_0 ^2$ units,
and in the second case, the values of $p^2$ are multiples of $m_0 ^2$.}
\label{Qpl}
\end{figure}

In Fig.\ref{Qpl} we display the behaviour of $Q$ for
selected values  of each parameter. From Fig.\ref{Qpl}(a) we
notice that for $0<\eta\leq 2$, $Q$ is always positive.
When $\eta>2$, $Q$ can become negative for high momenta. As $\eta$ grows,
$Q$ is positive as long as the spatial momentum is not so much greater
than $m_0$. This behaviour is confirmed in Fig.\ref{Qpl}(b).

In summary, the presence of the Lorentz violating mass terms in
the extended  Ho\u rava-Lifshitz theory, at quadratic level
without giving qualitative new results, extents the parametric
space of the theory giving more freedom for the constraints to be
satisfied in order not to have ghosts and instabilities.

\subsection{Strong Coupling Problem}

The strong coupling problem arises when quantum fluctuations start
to be important above a certain scale, and this can be understood
in the classical theory as the scale where the perturbation theory
breaks down. The strong coupling problem appears in theories which
have  a low energy (large distance) limit and it results to the
effect that the interactions become strong at a relatively low
energy scale. In the Ho\u rava-Lifshitz theory and in its
projectable version this problem was discussed in
\cite{Koyama:2009hc}.

Calculating the cubic interactions of the scalar graviton they
found that these interactions become important as $\l \to 1$, the
limit where in the IR General Relativity is recovered. The cubic
interactions are suppressed by the scale
$\Lambda=M_{Pl}c_{\zeta}^{3/2}$ which is relatively low scale
compared to the quadratic order. Then as $ c_{\zeta} \to 0 \,\,
(\lambda \to 1)$ the interactions blow up. They attributed this
pathology to the projectable condition.

A similar problem was discussed in \cite{Papazoglou:2009fj}. They
considered the extended Ho\u rava-Lifshitz theory of
\cite{Blas:2009qj}. They ignored the high derivative terms of the
three-vector field $\alpha_i$ of (\ref{ai}). They calculated the
cubic scalar interactions of the scalar graviton and they found
that they scale as $(\mid c_{\zeta}\mid M_{Pl})^{-1}$. Therefore
the cubic scalar interactions become significant as $ c_{\zeta}
\to 0 \,\, (\lambda \to 1)$ so there is a strong coupling problem
at the scale $\mid c_{\zeta}\mid M_{Pl}$.

It was pointed  out however in \cite{Blas:2009ck}, that this scale
is an IR energy scale and it is a result of a low energy
approximation. It was claimed that there is another UV scale
$\Lambda$ in the theory, suppressing the higher-derivative
operators which arise because of the presence of the three-vector
field, and if this UV scale is lower than the IR scale, then there
is no strong coupling problem.

This idea of introducing  the $\Lambda$ scale for suppressing the
high order operators, so that these operators could take over the
perturbation expansion before strong coupling kicks in, clearly
requires tuning and  arises naturalness issues in the sense that
we have to introduce a Lorentz violating scale much lower than the
Planck scale. Also there may be some consistency problems arising
from the observations, because the closer to 1 observations
required $\lambda$ to be, the smaller the scale $\Lambda$ should
be chosen to be as well. These results were also obtained and
further elaborated in \cite{arXiv:1003.5666} were the
St\"{u}ckelberg formalism was used.

Let us see how the presence   of the Lorentz violating masses
affects  the strong  coupling problem of the extended Ho\u
rava-Lifshitz theory. We substitute the canonically normalized
$\zeta$ field (\ref{normalized}) in the cubic Lagrangian
(\ref{cubiclag}). We are interested in the behaviour of the cubic
interactions in the limit $(\lambda \to 1)$. We find that the
strong coupling scale is at $ (m_1 M_{Pl})^{1/2}$. This scale at
which the interactions become strong is relatively large and
therefore we evade the problem of introducing ad hoc a low Lorentz
violating scale. Note here, that the same strong coupling scale
arises  in   the Lorenz violating theories of
\cite{Rubakov:2004eb,hep-th/0409124} and this is one of the main
motivations for introducing these theories in order to avoid the
strong coupling problem in massive gravity.

An interesting question is what happens with the mixing terms at
the cubic order, specially the mixing terms between tensor and
scalar perturbations. Do they introduce another scale? In the
Fierz--Pauli case, the energy scale that suppresses strong
interactions is $\Lambda=(m^4 M_{Pl})^{1/5}$ \cite{AHGS}. However,
if one considers the coupling of the scalar mode of the graviton
with an appropriate choice of higher order terms $h_{\mu \nu}$
(see a discussion on this in \cite{Rubakov:2008nh}) then the scale
is raised to $\Lambda=(m^2 M_{Pl})^{1/3}$. We have not calculated
the tensor perturbations in our case. We can argue however, that
even if another scale is generated through some kind of mixing
terms, in this scale some power of the mass  $m_1$ will appear,
making the strong coupling scale large enough.

\subsection{Absence of Tachyons}

To avoid tachyonic masses for the $\zeta$ field we analyse  the
dispersion relation (\ref{dispersion}). We have two possible
cases,
\begin{enumerate}

\item If $m_2 ^2 -3m_3 ^2 \geq 0$, then $\omega^2>0$ since $P$ and
$Q$ are already positive. However, this is not compatible with the
requirement to avoid a tachyon behaviour, {\it i.e.}, \be
\label{tachyon} (3m_3 ^2 -m_2 ^2) -3\frac{m_4 ^4}{m_0 ^2}>0~, \ee
so we must invoke  the second case.

\item If $m_2 ^2 -3m_3 ^2 < 0$, we need $Q>3(3m_3 ^2-m_2 ^2)$ to
ensure $\omega^2>0$. In this case, this possibility just needs
$$
3m_3 ^2 -m_2 ^2 > 3\frac{m_4 ^4}{m_0 ^2}
$$
in order to avoid a tachyon.

\end{enumerate}

\section{Conclusions}

We studied a Lorentz violating massive gravity theory. At short
distances the theory behaves like Ho\u rava: There exists  a
preferred foliation by three-dimensional constant time
hypersurfaces, which splits spacetime into space and time keeping
a three-dimensional Euclidean space invariance. This allows to add
 higher order spatial derivatives of the metric
to the action  improving the UV behaviour of the graviton
propagator and renders the theory power-counting renormalisable.
This theory in its most general form allows the introduction of a
three-dimensional vector invariant under a three-dimensional
Euclidean symmetry. The presence of this vector and the higher
order operators constructed out of it, are supposed to improve the
behaviour of the theory in the IR.

 Since this theory is Lorentz
violating even at large distances, it is natural to introduce
Lorentz violating mass terms. We know that the presence of these
terms makes the massive gravity theory healthy at large distances
\cite{Rubakov:2004eb,Rubakov:2008nh}: no ghosts, the vDVZ
discontinuity  is absent, and there is no strong coupling problem.
These terms in short distances appear as dynamical metric fields
with dimensionful running coupling constants which do not spoil
the power-counting renormalisability of the theory. We expect that
the running of these dimensionful coupling constants through some
RG flow to give terms of dimension mass-squared at the IR.

We made a detailed study of this theory at large distances. By
considering scalar perturbations we show that for a wide range of
parameters, the scalar mode of the graviton has healthy behaviour:
it has normal kinetic term, it is not tachyonic and it does not
suffer from  strong coupling problems and instabilities.

We did not study  tensor and vector perturbations of the theory
restricting ourselves to the  scalar perturbations which are known
to have unwanted pathologies like ghosts and instabilities. For
the tensor perturbations we expect that the results of
\cite{Rubakov:2004eb,Rubakov:2008nh} will be reproduced, while
 the presence of the three-vector field $a_i$ will contribute to the
 analysis  for the vector
perturbations. This issue merits further study.

\section*{Acknowledgments}

We thank  Elcio Abdalla, Alex Kehagias, Elias Kiritsis, and
Antonis Papazoglou for constructive comments and remarks. B.C-M.
acknowledges partial support of the State Scholarships Foundation
(IKY) under contract 1288 and also the support of Consejo Nacional
de Investigaciones Cienti'ficas y Te'cnicas (CONICET).

\end{document}